\documentclass[preprint,english,aps,prl]{revtex4}%
\usepackage{amsfonts}
\usepackage[T1]{fontenc}
\usepackage[latin9]{inputenc}
\usepackage{amsmath}
\usepackage{amssymb}
\usepackage{babel}
\usepackage{babel}
\usepackage{babel}
\usepackage{graphicx}%
\makeatletter
\@ifundefined{textcolor}{}
{
\definecolor{BLACK}{gray}{0}
\definecolor{WHITE}{gray}{1}
\definecolor{RED}{rgb}{1,0,0}
\definecolor{GREEN}{rgb}{0,1,0}
\definecolor{BLUE}{rgb}{0,0,1}
\definecolor{CYAN}{cmyk}{1,0,0,0}
\definecolor{MAGENTA}{cmyk}{0,1,0,0}
\definecolor{YELLOW}{cmyk}{0,0,1,0}
}
\makeatother
\begin{document}
\title{Weak values from path integrals}
\author{A. Matzkin}
\affiliation{Laboratoire de Physique Théorique et Modélisation, CNRS Unité 8089,
CY Cergy Paris Université, 95302 Cergy-Pontoise cedex, France}

\begin{abstract}
We connect the weak measurements framework to the path integral formulation of
quantum mechanics. We show how Feynman propagators can in principle be
experimentally inferred from weak value measurements. We also obtain
expressions for weak values parsing unambiguously the quantum and the
classical aspects of weak couplings between a system and a probe. These
expressions are shown to be useful in quantum chaos related studies (an
illustration involving quantum scars is given), and also in solving current
weak-value related controversies (we discuss the existence of discontinuous
trajectories in interferometers and the issue of anomalous weak values in the
classical limit). 

\end{abstract}
\maketitle

There has been a growing interest in weak measurements -- a specific form of
quantum non-demolition measurements -- over the last decade. Weak measurements
were indeed found to be useful in fundamental or technical investigations,
involving both experimental and theoretical works. Nevertheless, ever since
their inception \cite{AAV}, weak values -- the outcomes of weak measurements
-- have remained controversial. Since weak values are entirely derived from
within the standard quantum formalism, the controversies have never concerned
the validity of weak values (WV), but their understanding and their
properties.\ For instance, are WV values similar to eigenvalues, or are they
akin to expectation values \cite{vaidmanW,dresselRMP}? Do anomalous WV
represent a specific signature of quantum phenomena or can they be reproduced
by classical conditional probabilities \cite{combes,pusey,dressel,jordan}? Are
WV\ related to the measured system properties or do they represent arbitrary
numbers characterizing the perturbation of the weakly coupled pointer
\cite{matzkin-found,sokolovski2016}?

The path integral formulation of quantum mechanics is strictly equivalent to
the standard formalism based on the Schrödinger equation.\ Though it is often
technically more involved, path integrals give a conceptually clearer picture
though the natural, built-in connection with quantities defined from classical
mechanics (Lagrangian, action, paths).\ Surprisingly, very few works have
employed weak values in a path integral context. Even then, the interest was
restricted to the weak measurement of Feynman paths in semiclassical systems
\cite{matzkinPRL,tsutsuiPTP,Narducci}, to WV of specific operators
\cite{tanaka,hiley,sokolovskiFP}, or as a way to probe virtual histories
\cite{cohenPRA}.

In this work, we connect the weak measurements framework to the path integral
formulation. We will see that path integrals parse the quantum and classical
aspects of weak values, thereby clarifying many of the current controversies
involving weak measurements. In particular we will show how the path integral
expression accounts for the discontinuous trajectories observed in
interferometers and currently wildly debated \cite{past-exp,pastQP,past-int}.
We will also see that in the classical limit the weak value is washed out by
coarse graining, implying that anomalous pointer values are a specific quantum
feature with no classical equivalent \cite{combes,pusey,dressel}. A nice
feature arising from the present approach is the possibility to measure
propagators through weak measurements. This could be a useful tool in quantum
chaos related studies as a method to observe the semiclassical amplitudes; we
give an illustration in which the autocorrelation function employed to
investigate quantum scars in inferred from weak values. The present approach
could also be the starting point to apply weak values in domains, such as
quantum cosmology, for which the standard, non-relativistic weak measurements
framework, relying on a von Neumann interaction, cannot be applied.

A weak measurement of a system observable $\hat{A}$ is characterized by 4
steps: 1. The system of interest is prepared in the chosen initial state
$\left\vert \psi_{i}\right\rangle $, a step known as preselection. A quantum
probe is prepared in a state $\left\vert \phi_{i}\right\rangle ,$ and the
initial state is thus
\begin{equation}
\left\vert \Psi(t_{i})\right\rangle =\left\vert \psi_{i}\right\rangle
\left\vert \phi_{i}\right\rangle . \label{initst}%
\end{equation}
2. The system and the probe are weakly coupled through an interaction
Hamiltonian $\hat{H}_{int}$ (a so-called von Neumann interaction). 3. After
the interaction, the system evolves until another system observable, say
$\hat{B}$, is measured through a standard measurement process.\ Of all the
possible eigenstates $\left\vert b_{k}\right\rangle $ that can be obtained, a
filter selects only the cases for which $\left\vert b_{k}\right\rangle
=\left\vert b_{f}\right\rangle $, where $\left\vert b_{f}\right\rangle $ is
known as the postselected state. 4. When postselection is successful, the
probe is measured.\ The final state of the probe $\left\vert \phi
_{f}\right\rangle $ has changed relative to $\left\vert \phi_{i}\right\rangle
$ by a shift depending on the weak value of $\hat{A}$.

Let $\hat{U}_{s}(t_{2},t_{1})$ denote the system evolution operator, and let
$t_{i},t_{w}$ and $t_{f}$ represent the preparation, interaction, and
postselection times resp. The weak value \cite{AAV} of $\hat{A}$ is then given
by
\begin{equation}
A^{w}=\frac{\left\langle b_{f}(t_{f})\right\vert \hat{U}_{s}(t_{f},t_{w}%
	)\hat{A}\hat{U}_{s}(t_{w},t_{i})\left\vert \psi(t_{i})\right\rangle
}{\left\langle b_{f}(t_{f})\right\vert \hat{U}_{s}(t_{f},t_{i})\left\vert
	\psi(t_{i})\right\rangle }. \label{awd}%
\end{equation}
Typically the probe state $\left\vert \phi_{i}\right\rangle $ is a Gaussian
pointer initially centered at the position $Q_{w}$ and $\hat{H}_{int}$ is of
the form
\begin{equation}
\hat{H}_{int}=g(t)\hat{A}\hat{P}f_{w} \label{hint}%
\end{equation}
where $\hat{P}$ is the probe momentum, $g(t)$ is a function non-vanishing only
in a small interval centered on $t_{w}$, and $f_{w}$ reflects the short-range
character of the interaction that is only non-vanishing in a small region near
$Q_{w}.$\ Under these conditions it is well-known \cite{AAV} that $\left\vert
\phi_{f}\right\rangle =e^{-igA^{w}\hat{P}}\left\vert \phi_{i}\right\rangle ,$
with $g=\int dt^{\prime}g(t^{\prime})$ and the final probe state is the
initial Gaussian shifted by $g\operatorname{Re}(A^{w}).$

In order to determine the evolution of the coupled system-probe problem from
the initial state $\left\vert \Psi_{i}(t_{i})\right\rangle =\left\vert
\psi_{i}\right\rangle \left\vert \phi_{i}\right\rangle $, we need the full
Hamiltonian $\hat{H}=\hat{H}_{s}+\hat{H}_{p}+\hat{H}_{int}$ where $\hat{H}%
_{s}$ and $\hat{H}_{p}$ are the Hamiltonians for the uncoupled system and
probe resp. In terms of the corresponding evolution operators, the system and
the probe evolve first independently, $\hat{U}_{0}(t,t_{i})=\hat{U}%
_{s}(t,t_{i})\hat{U}_{p}(t,t_{i}).$ Then, assuming for simplicity that the
interaction takes place during the time interval $\left[  t_{w}-\tau
/2,t_{w}+\tau/2\right]  $ ($\tau$ is the duration of the interaction), the
total evolution operator from $t_{i}$ to the postselection time $t_{f}$ is
given as
\begin{align}
\hat{U}(t_{f},t_{i})=\hat{U}_{0}(t_{f},t_{w}+\tau/2)\hat{U}_{int}  &
(t_{w}+\tau/2,t_{w}-\tau/2)\nonumber\\
&  \hat{U}_{0}(t_{w}-\tau/2,t_{i}). \label{unitary}%
\end{align}
The propagators $K\equiv\left\langle x_{2}\right\vert \hat{U}(t_{2}%
,t_{1})\left\vert x_{1}\right\rangle $ for the uncoupled evolution of the
system and probe are given resp. by
\begin{align}
&  K_{s}(x_{2},t_{2};x_{1},t_{1})=\int_{x_{1}}^{x_{2}}\mathcal{D}%
[q(t)]\exp\left(  \frac{i}{\hbar}\int_{t_{1}}^{t_{2}}L_{s}(q,\dot{q}%
,t^{\prime})dt^{\prime}\right) \label{propp}\\
&  K_{p}(X_{2},t_{2};X_{1},t_{1})=\int_{X_{1}}^{X_{2}}\mathcal{D}%
[Q(t)]\exp\left(  \frac{i}{\hbar}\int_{t_{1}}^{t_{2}}L_{p}(Q,\dot
{Q},t)dt^{\prime}\right)  \label{props}%
\end{align}
where as usual \cite{schulman,Feynman-Hibbs} $\mathcal{D}[.]$ implies
integration over all paths connecting the initial and final space-time points,
and $L_{s}=\frac{m\dot{q}^{2}}{2}-V(q)$ and $L_{p}=\frac{M\dot{Q}^{2}}{2}$ are
the classical system and probe Lagrangians resp.

For the coupled evolution $\hat{U}_{int},$ the propagator becomes
non-separable,
\begin{align}
K_{int}(X_{2},x_{2},t_{2};X_{1},x_{1},t_{1})  &  =\int_{\left(  X_{1}%
	,x_{1}\right)  }^{\left(  X_{2},x_{2}\right)  }\mathcal{D}[Q(t)]\mathcal{D}%
[q(t)],\nonumber\\
&  \times\exp\left[  \frac{i}{\hbar}\int_{t_{1}}^{t_{2}}L\left(  Q,\dot
{Q},q,\dot{q},t^{\prime}\right)  dt^{\prime}\right]  \label{intt}%
\end{align}
where
\begin{equation}
L=L_{s}+L_{p}-g(t)A(q)f(q,Q_{w})M\dot{Q} \label{lag}%
\end{equation}
is the classical interacting Lagrangian \cite{SM}; $f(q,Q_{w})$ sets
the range of the interaction (it becomes a Dirac delta function in the limit
of a point-like interaction).\ $Q_{w}$ will be taken here as a parameter
specifying the position of the probe, which makes sense for a probe with a
negligible kinetic term. Note that $A(q)$ gives the configuration space value
of the classical dynamical variable $A(q).$

Non-separable propagators are notoriously difficult to handle except when they
can be treated perturbatively \cite{Feynman-Hibbs}, which is the case here.
Standard path integral perturbation techniques (see Ch. 6 of
\cite{Feynman-Hibbs}) applied to the system degrees of freedom give $K_{int}$
in terms of the uncoupled propagators and a first order correction in which
the system propagates as if it were uncoupled except that each path $q(t)$ is
weighed by the perturbative term $g(t)A(q)f(q,Q_{w})M\dot{Q}$. If the duration
$\tau$ of the interaction is small relative to the other timescales (as is
usually assumed in weak measurements), the time-dependent coupling can be
integrated to an effective coupling constant $g=\int_{t_{w}-\tau/2}%
^{t_{w}+\tau/2}g(t^{\prime})dt^{\prime}$ and the uncoupled paths see an
effective perturbation $gA(q)f(q,Q_{w})M\dot{Q}$ at time $t_{w}$. Eq.
(\ref{intt}) becomes (the derivation is given in the Supp.\ Mat. \cite{SM})
\begin{align}
K_{int}  &  =K_{p}K_{s}+\int_{X_{1}}^{X_{2}}\mathcal{D}[Q(t)]e^{\frac{i}%
	{\hbar}\int_{t_{1}}^{t_{2}}L_{p}dt^{\prime}}\int dqK_{s}(x_{2},t_{2}%
;q,t_{w})\nonumber\\
&  \times A(q)f(q,Q_{w})K_{s}(q,t_{w};x_{1},t_{1})\left(  -\frac{ig}{\hbar
}M\dot{Q}\right)  . \label{intts}%
\end{align}

From the point of view of the system, the interpretation of Eq. (\ref{intts})
is straightforward (see Fig. 1): the transition amplitude from $x_{i}$ to
$x_{f}$ involves a sum over the paths directly joining these 2 points in time
$t_{f}-t_{i}$ as well as those that interact with the probe in the region
determined by $f(q,Q_{w})$, hence going from $x_{i}$ to some intermediate
point $q$ within this region, and then from this point $q$ to $x_{f}$.

We now take into account the pre and postselected states and focus on the
probe evolution. For convenience we perform the propagation from the initial
state $\left\vert \Psi_{i}\right\rangle =\int dX_{i}dx_{i}\left\vert
X_{i}\right\rangle \left\vert x_{i}\right\rangle \psi(x_{i})\phi(X_{i})$ up to
the interaction time $t_{w}$. The postselected state $\left\langle
b_{f}\right\vert =\int dxb_{f}^{\ast}(x_{f})\left\langle x_{f}\right\vert $ is
instead propagated backwards to $t_{w}$. The uncoupled evolution $\left\langle
b_{f}(t_{w})\right\vert \left.  \psi_{i}(t_{w})\right\rangle \phi_{i}%
(X,t_{f})$ involves the direct paths from each $x_{i}$ where $\psi$ is
non-vanishing to each $x_{f}$ lying in the support of $b_{f}(x)$. Factorising
this term in the full evolution, we get \cite{SM}
\begin{align}
\phi(X,t_{f})  &  =\left\langle b_{f}(t_{w})\right\vert \left.  \psi_{i}%
(t_{w})\right\rangle \nonumber\\
&  \int dX_{1}\int_{X_{1}}^{X}D[Q(t)]e^{\frac{i}{\hbar}\left[  \int_{t_{w}%
	}^{t_{f}}L_{p}dt^{\prime}-gA_{w}M\dot{Q}\right]  }\phi_{i}(X_{1},t_{w})
\label{probp}%
\end{align}
where $A^{w}$ is the weak value of $\hat{A}$ given by\begin{widetext}
	\begin{equation}
	A^{w}=\frac{\int A(q)f(q,Q_{w})K_{s}(x_{f},t_{f};q,t_{w})K_{s}(q,t_{w};x_{i},t_{i})b_{f}^{\ast}(x_{f})\psi(x_{i})dqdx_{f}dx_{i}}{\int dx_{f}dx_{i}K_{s}(x_{f},t_{f};x_{i},t_{i})b_{f}^{\ast}(x_{f})\psi(x_{i})}.\label{wve}
	\end{equation}
\end{widetext}

Several remarks are in order. First note that for large $M$ Eq. (\ref{probp})
implies a shift of the initial probe state, since the Lagrangian maps each
point $X_{1}$ to $X_{1}-gA_{w}$. Second, while the denominator in Eq.
(\ref{wve}) involves all the paths connecting the initial region to the final
region, the numerator contains the sole paths connecting the initial and final
regions passing through a point $q$ of the interaction region. Each such path
is weighed by the \emph{classical value} of the configuration space function
$A(q)$ at that point.\ Note also that the weak value expressions (\ref{awd})
and (\ref{wve}) are both the results of asymptotic expansions,but the
expansions are not exactly equivalent. For a contact interaction
$f(q,Q_{w})=\delta(q-Q_{w})$, it is easy to see that Eqs. (\ref{awd}) and
(\ref{wve}) become identical under the condition $A(q)=\left\langle
q\right\vert \hat{A}\left\vert q\right\rangle $ for $q=Q_{w}$. In this case
$A^{w}$ is simply given by $A(Q_{w})\times T_{w}/T$ where $T_{w}$ is the
transition amplitude involving the paths connecting $x_{i}$ to $x_{f}$ by
going through $Q_{w}$ at $t_{w}$, while $T$ connects $x_{i}$ to $x_{f}$
through any intermediate point) (see Fig. 1 and Eq. (\ref{wvsimp}) below). We
then see that: (i) The ratio $T_{w}/T$ can take any complex value, hence if
$A(Q_{w})$ is bounded, $\Re(A^{w})$ can lie beyond these bounds and $A^{w}$ is
said to be \textquotedblleft anomalous\textquotedblright\ \footnote{The
	standard definition of anomalous -- $\Re(A^{w})$ can lie above the largest or
	below the lowest eigenvalue -- is more restrictive since by canonical
	quantization the eigenvalues range is typically contained within the possible
	values of $A(q)$ in configuration space.}. (ii) If the postselection state is
chosen to be the space point $\left\vert x_{f}\right\rangle ,$ we have
$A^{w}=A(Q_{w})K_{s}(x_{f},Q_{w})\psi(Q_{w},t_{w})/\psi(x_{f},t_{f}),$ so if
the wavefunctions are known, eg through a previous weak measurement based
procedure \cite{lundeen}, the propagator $K_{s}(x_{f},Q_{w})$ can be obtained
from the measurement of the weak value $A^{w}$. Both $x_{f}$ and $Q_{w}$ can
be varied in order to measure the propagator over the region of interest.

\begin{figure}[t]
	\centering \includegraphics[scale=0.27]{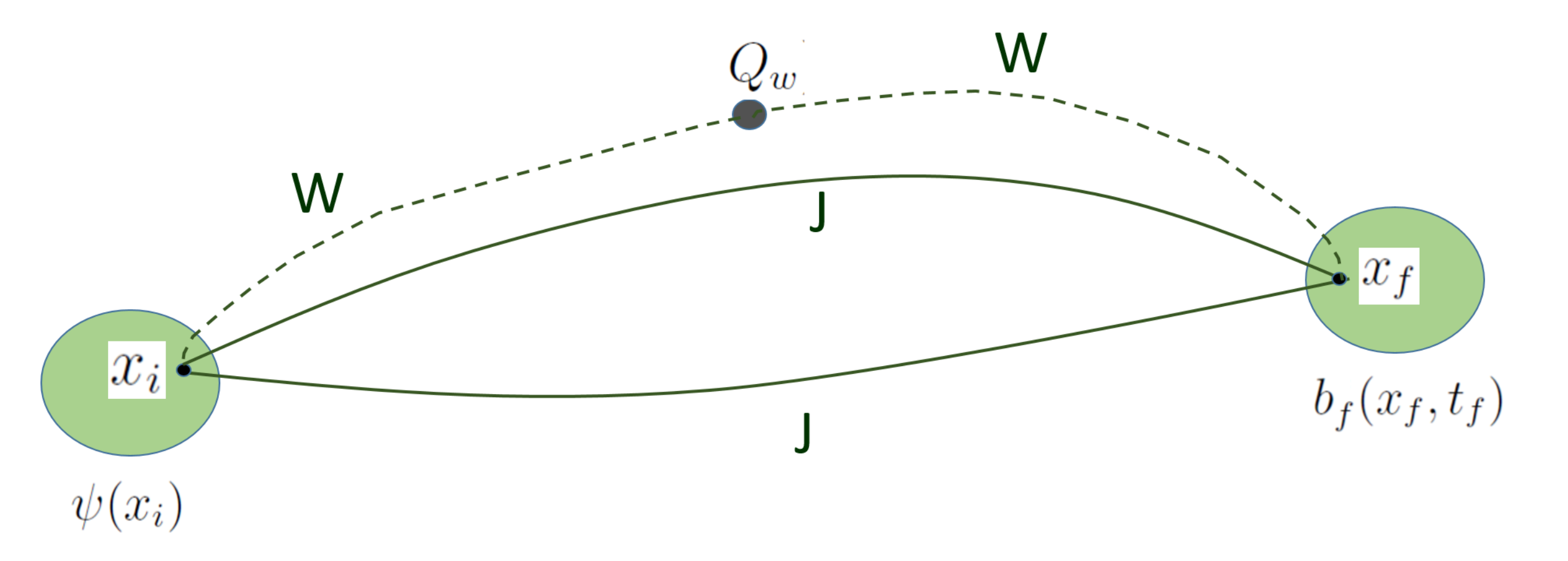} \label{fentes}%
	\caption{Schematic illustration of the paths involved in the weak value
		expression when the pre- and postselected states are well localized, and the
		propagator is given as a sum over the classical trajectories [see Eq.
		(\ref{wvsimp})]. For the specific choice of postselection $b(x_{f}%
		)=\delta(x-x_{f})$ the weak value resulting from the coupling at $q=Q_{w}$
		becomes proportional to the propagator $K_{s}(x_{f},Q_{w})$. The propagator
		can hence be observed from weak value measurements varying $Q_{w}$ and $x_{f}%
		$.}%
\end{figure}

Our weak value expression (\ref{wve}) is useful in cases involving free
propagation or in the semiclassical approximation: for large actions
stationary phase integration transforms the sum over all arbitrary paths to a
sum containing only the classical paths linking the initial and final points,
so that both propagators (\ref{propp}) and (\ref{props}) take the form
$K^{sc}=\sum_{k}\mathcal{A}_{k}\exp i\left[  \mathcal{S}_{k}/\hbar-\pi\mu
_{k}/2\right]  $ where $\mathcal{A}_{k}$ and $\mathcal{S}_{k}$ are the
amplitude and classical action for the classical path $k$ connecting the
initial and endpoints in time $t_{f}-t_{i}$ \cite{schulman,brack}; the phase
index $\mu_{k}$ counts the number of conjugate points along each trajectory
and will be absorbed into the action to simplify the notation. The
semiclassical regime, exact for free propagation, remains quantum since the
different classical paths still interfere, and $A^{w}$ can be anomalous.
Assuming again a point-like interaction at $q=Q_{w},$ Eq. (\ref{wve}) becomes
\begin{widetext}
	\begin{equation}
	A^{w}=A(Q_{w})\frac{\int dx_{f}dx_{i}\chi_{f}^{\ast}(x_{f})\psi\left(x_{i}\right)\left[\sum_{W}\mathcal{A}_{W}(x_{f};Q_{w})e^{i\mathcal{S}_{W}(x_{f};Q_{w})/\hbar}\mathcal{A}_{W}(Q_{w};x_{i})e^{i\mathcal{S}_{W}(Q_{w};x_{i})/\hbar}\right]}{\int dx_{f}dx_{i}\chi_{f}^{\ast}(x_{f})\psi\left(x_{i},t_{i}\right)\sum_{J}\mathcal{A}_{J}(x_{f};x_{i})e^{i\mathcal{S}_{J}(x_{f};x_{i})/\hbar}}.\label{wvsimp}
	\end{equation}
\end{widetext}$\mathcal{A}_{W}$ and $\mathcal{S}_{W}$ label the amplitude and
action of a path going through $Q_{w}$, $W$ runs over all the classical paths
connecting $x_{i}$ to $Q_{w}$ and $Q_{w}$ to $x_{f},$ while $J$ runs over all
the classical paths connecting directly $x_{i}$ to $x_{f}$ in time
$t_{f}-t_{i}.$ \ 

The form of the weak value given by Eq. (\ref{wvsimp}) is particularly
well-suited to the experimental investigation of the contribution of
individual orbits in semiclassical systems, such as the detection of quantum
scars in wavefunctions that have recently received renewed interest \cite{scars}. The standard
approach \cite{heller84} to periodic orbit localization involves the
autocorrelation function $\left\langle G(0)\right\vert \left.
G(t)\right\rangle $ where $G(x,0)$ is a Gaussian with a maximum $x_{0}$ placed
on a periodic orbit (po) that quickly spreads over all the available
phase-space. If the system is prepared and post-selected in state $\left\vert
G(0)\right\rangle $ the weak value of any observable $\hat{A}$ measured on a
point $x_{p}$ of the periodic orbit is given by $A^{w}=\left\langle
G(0)\right\vert \hat{U}_{s}(t_{f},t_{w})\Pi_{x_{p}}\hat{A}\Pi_{x_{p}}\hat
{U}_{s}(t_{w},t_{i})\left\vert G(0)\right\rangle /\left\langle G(0)\right\vert
\left.  G(t_{f})\right\rangle $ where $\Pi_{x_{p}}\equiv\left\vert
x_{x_{p}}\right\rangle \left\langle x_{x_{p}}\right\vert $ is the projector on  $x_{p}$. By using Eq. (\ref{wvsimp}), the
autocorrelation function is obtained from the knowledge of the (presumably
observed) weak value and of a single Feynman path, the po :
\begin{align}
\left\langle G(0)\right\vert  &  \left.  G(t_{f})\right\rangle =A^{w}%
\times\label{autoc}\\
&  \left(  \mathcal{A}_{po}(x_{0},x_{p})A(x_{p})\mathcal{A}_{po}(x_{p}%
,x_{0})e^{iS_{po}/\hbar}\left\vert G(x_{0},0)\right\vert ^2\right)
^{-1}.\nonumber
\end{align}
Instead, the application of the standard weak value definition (\ref{awd}) to
infer the autocorelation function would require the knowledge of the full
Schrödinger propagator.

\begin{figure}[t]
	\centering \includegraphics[scale=0.27]{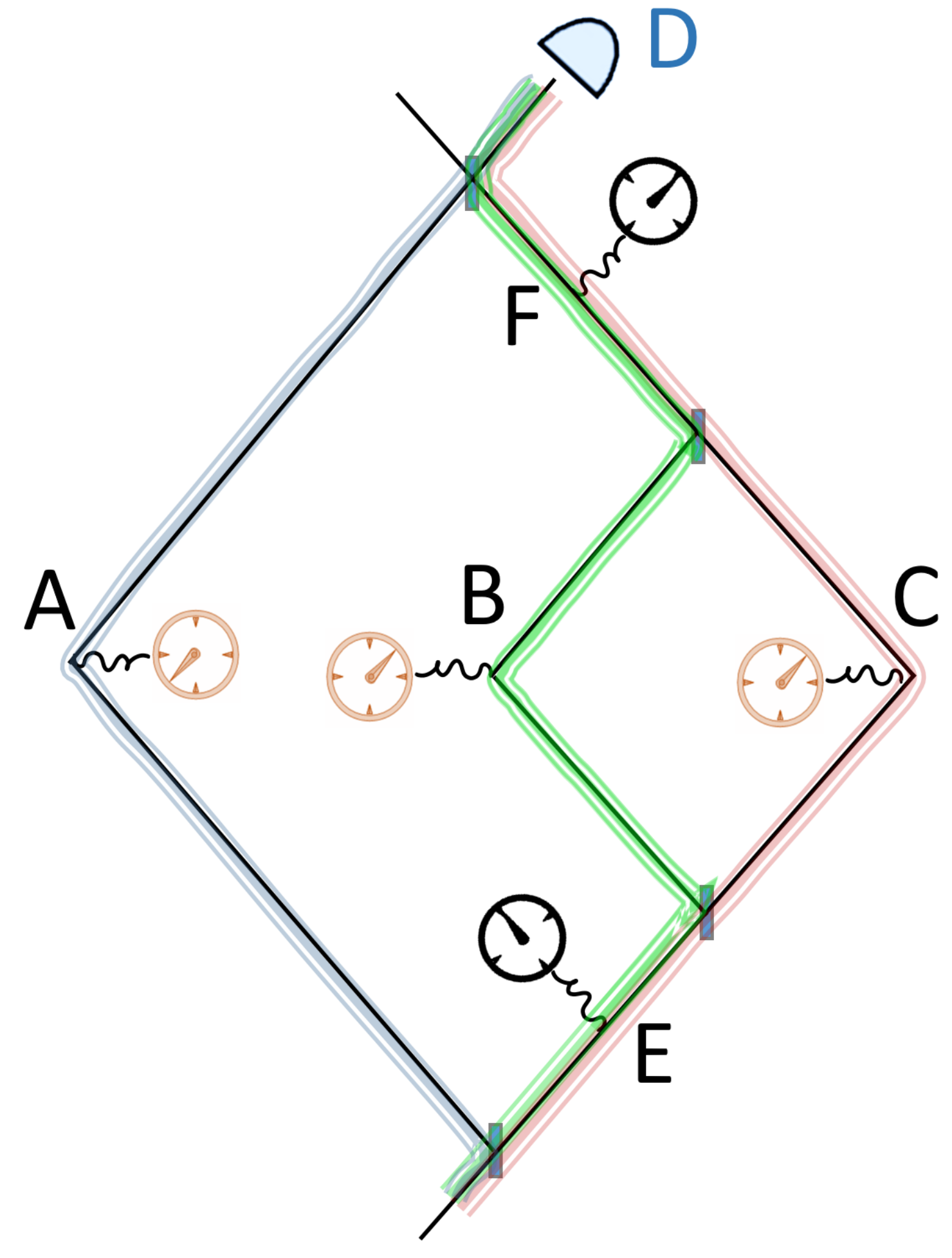} \caption{A quantum
		particle propagating in an interferometer and postselected at $D$ displays
		discontinuous trajectories when observed with weak measurements (the weakly
		coupled probes depicted in black remain unaffected by the interaction with the
		particle, while the probes depicted in orange see their pointers shift after
		postselection, see Eq. (\ref{pqp})). The Feynman paths of the particle,
		represented as pencils of trajectories along the arms, are continuous and
		progagate the interactions with the probes up to $D$, where superposition with
		the postselected state yield zero WV at $E$ and $F$ even when the wavefunction
		on these segments does not vanish (see text for details).}%
	\label{paths-nMZ}%
\end{figure}

Our approach is also relevant to understand current controversies involving WV
such as the apparent observation of discontinuous trajectories
\cite{past-exp,pastQP,past-int} or the quantumness of anomalous weak values
\cite{combes,pusey,dressel}, as we briefly detail. The first of these issues
involves a 3-paths interferometer depicted in Fig.\ \ref{paths-nMZ}.\ The
weakly measured observable is the projector $\Pi_{j}\equiv\left\vert
x_{j}\right\rangle \left\langle x_{j}\right\vert $ indicating whether the
particle is at point $x_{j}$.\ By suitably choosing the preselected and
postselected states, the following weak values are obtained \cite{past-exp}:
\begin{equation}
\Pi_{A}\neq0,\Pi_{E}=0,\Pi_{B}\neq0,\Pi_{C}\neq0,\Pi_{F}=0.\label{pqp}%
\end{equation}
This means that weakly coupling a probe to the system leaves a trace on paths
$A,$ $B$ and $C,$ but none on the segments $E\ $and $F$: the particle is seen
inside the loop, but not on the entrance and exit paths. A point that has
caused some confusion in the literature \cite{pastQP,past-int} is that in the
initial proposal \cite{past-exp} the wavefunction vanished (by destructive
interference) on paths $E$\ and $F$, so that an interpretation in terms of
vanishing weak values appeared to be flawed. It is possible however to enforce
Eq. (\ref{pqp}) without having destructive interference on $E$ and $F$. The
present path integral approach naturally parses the aspects relevant to a
vanishing classical value, to a vanishing superposition, or to a partially
propagated state being incompatible with postselection. Indeed, the WV
expression (\ref{wvsimp}) vanishes

\begin{enumerate}
	\item[(i)] if $A(Q_{w})=0$, which for a projector implies that the particle is
	not there;
	
	\item[(ii)] if the term between brackets vanishes, corresponding to
	destructive interference of the wavefunction at $Q_{w}$ by summing over the
	different continuous paths $W$;
	
	\item[(iii)] if the integral vanishes, that is the ensemble of points of the
	preselected state \emph{propagated by the sole paths} going through $Q_{w}$ up
	to $x_{f}$ is orthogonal to the postselected state.
\end{enumerate}

The case for asserting that the particle is not at $Q_{w}=E$ or $F$ when the
WV vanishes is unproblematic in case (i): this is independent of postselection
and would also be the case classically. (ii) corresponds to the initial
proposal \cite{past-exp} in which 2 paths with opposite phases interfere
destructively.\ This is a non-classical effect (since we have $A(Q_{w})\neq0$)
but it does not necessarily depend on postselection and an interpretation in
terms of weak values appears moot: if there is no wavefunction, there is
nothing to measure, as remarked in Refs \cite{past-int}. The genuinely
interesting case is (iii): the paths do not interfere destructively in the interaction
region, but the fraction of the 
preselected wavefunction propagated by the paths passing through this region (here $E$ or
$F$) is incompatible with the postselected state. In other interaction regions (here $B$ or $C$) the
paths contributing to the propagator in Eq. (\ref{wvsimp}) are not orthogonal to the
postselected state anymore. The crucial point is that the
paths contained in the propagator are continuous, but each probe's motion
results from the overlap at $t=t_{f}$ between the preselected state propagated
by the paths hitting the probe region and the postselected state.

The other current controversy mentioned above concerns the quantumness of
anomalous weak values that has been recently questioned as resulting from a
mere statistical effect \cite{combes,pusey,dressel,jordan}. A path integral
approach is convenient to obtain heuristically the classical limit of the
weakly coupled probe's motion.\ This involves two well-known steps. The first
step, deriving the semiclassical propagator $K^{sc}$ was recalled above. The
second one involves retaining only the diagonal terms of the semiclassically
propagated density matrix \cite{bhatt}, due to coarse graining over a
classical scale \cite{classlim}. We then obtain \cite{SM} classical
amplitudes obeying the classical transport equation, and the weak probe's
dynamics is given by solving for each point of the initial distribution the
equations of motion for the Lagrangian given by Eq. (\ref{lag}). Postselection
is specified in terms of a domain $\mathfrak{B}_{f}$ at time $t_{f}$ defined
such that the chosen configuration space function $b(q,t_{f})\in
\mathfrak{B}_{f}$. The average probe shift with postselection is then%
\begin{equation}
\left\langle \Delta Q\right\rangle =g\int_{\mathfrak{B}_{w}}A(q,t_{w}%
)f(q,Q_{w})\frac{\rho_{s}(q,t_{w})}{\int_{\mathfrak{B}_{w}}\rho_{s}(q^{\prime
	},t_{w})dq^{\prime}}dq.\label{classps}%
\end{equation}
The integral is taken at $t_{w}$ over the configuration space domain
$\mathfrak{B}_{w}$ such that at $t_{f}$ the condition $b(q,t_{f}%
)\in\mathfrak{B}_{f}$ is obeyed. $\left\langle \Delta Q\right\rangle $ can
never be anomalous.\ Classically, the only way to have an anomalous probe
shift would be to replace $A(q,t_{w})$ by a different configuration space
function or to have the numerator and denominator integrated over different
domains, say $\mathfrak{B}_{f}$ and $\mathfrak{B}_{f}^{\prime}$. In both cases
this would be the result of a perturbation, due to the detection process in
the latter case.

To sum up, we have obtained an expression of weak values from a path integral
approach. We have shown this expression to be useful when semiclassical
propagators are involved. We have seen that the present approach gives a
consistent account of two current controversies involving weak values (the
observation of discontinuous trajectories and the quantumness of anomalous
weak values). Other recent perplexing results \cite{sense} involving pre and
post-selected ensembles can be treated similarly. \ We have also suggested a
method for measuring the Feynman propagator through weak values. The approach
introduced in this work will be fruitful to tackle extensions of the weak
measurements framework to relativistic and cosmological settings, where
quantities analogous to weak values are known to emerge \cite{englert}.

\renewcommand{\theequation}{A-\arabic{equation}}
\setcounter{equation}{0}

\section*{Appendix 1 - Classical Interacting Lagrangian and Propagator}

The full propagator (including the interaction) is given by Eq. (7) of the
text,%
\begin{align}
K_{int}(X_{2},x_{2},t_{2};X_{1},x_{1},t_{1})  &  =\int_{\left(  X_{1}%
	,x_{1}\right)  }^{\left(  X_{2},x_{2}\right)  }\mathcal{D}[Q(t)]\mathcal{D}%
[q(t)],\nonumber\\
&  \times\exp\left[  \frac{i}{\hbar}\int_{t_{1}}^{t_{2}}L\left(  Q,\dot
{Q},q,\dot{q},t^{\prime}\right)  dt^{\prime}\right] \label{Apropint}%
\end{align}
with the Lagrangian%
\begin{align}
L  &  =\frac{M\dot{Q}^{2}}{2}+\frac{m\dot{q}^{2}}{2}-V(q)-g(t)A(q)f(q,Q_{w}%
)M\dot{Q}\label{Alag}\\
&  \equiv L_{p}\qquad+\qquad L_{s}\qquad+\quad\quad L_{coupl}%
\end{align}
The classical equations of motion are
\begin{align}
\frac{d}{dt}\partial_{\dot{q}}L  &  =m\ddot{q}=\partial_{q}L=-\partial
_{q}V-g(t)M\dot{Q}\partial_{q}\left(  Af(q,Q_{w})\right) \\
\frac{d}{dt}\partial_{\dot{Q}}L  &  =M\ddot{Q}-\frac{d}{dt}g(t)Af(q,Q_{w}%
)M=\partial_{Q}L=0
\end{align}
and this implies that
\begin{align}
\dot{Q}-g(t)A\delta(q,q_{w})=cst\equiv P_{0}/M  & \\
\Delta Q\equiv Q(t>t_{w}+\tau/2)-Q_{0}=  &  \int_{t_{w}-\tau/2}^{t_{w}+\tau
	/2}g(t^{\prime})A(q)f(q,Q_{w})dt^{\prime}+\tau P_{0}/M\\
=  &  gA(q)f(q,Q_{w})\label{Ag}%
\end{align}
where the last line is obtained by neglecting $\tau P_{0}/M$ for large mass
probes and we have set
\begin{equation}
g\equiv\int_{t_{w}-\tau/2}^{t_{w}+\tau/2}g(t^{\prime})dt^{\prime}\label{Agdef}%
\end{equation}
as in the main text. We have set here $t_{1}=t_{w}-\tau/2$ and $t_{2}%
=t_{w}+\tau/2$ where $t_{w}$ is the mean interaction time, since for other
times only uncoupled propagation takes place.

The propagator (\ref{Apropint}) is computed in the weak coupling limit by
employing standard perturbation methods for path integrals (see Ch.\ 6 of
\cite{Feynman-Hibbs}). We assume the action term corresponding to the
interaction $\int_{t_{1}}^{t_{2}} L_{coupl} dt^{\prime}$ is small and can be
expanded to first order. Eq. (\ref{Apropint}) becomes%
\begin{align}
K_{int}(X_{2},x_{2},t_{2};X_{1},x_{1},t_{1})  &  =\int_{\left(  X_{1}%
	,x_{1}\right)  }^{\left(  X_{2},x_{2}\right)  }\mathcal{D}[Q(t)]\mathcal{D}%
[q(t)]\exp\left[  \frac{i}{\hbar}\int_{t_{1}}^{t_{2}}\left(  L_{s}%
+L_{p}\right)  dt^{\prime}\right] \nonumber\\
&  \times\left(  1-\frac{i}{\hbar}\int_{t_{1}}^{t_{2}}dt^{\prime}g(t^{\prime
})A(q)f(q,Q_{w})M\dot{Q}\right)  .
\end{align}
The time and path integrations in the perturbative term are interchanged,
yielding%
\begin{equation}
-\frac{i}{\hbar}\int_{t_{1}}^{t_{2}}dt^{\prime\prime}\int_{\left(  X_{1}%
	,x_{1}\right)  }^{\left(  X_{2},x_{2}\right)  }\mathcal{D}[Q(t)]\mathcal{D}%
[q(t)]e^{\frac{i}{\hbar}\int_{t_{1}}^{t_{2}}\left(  L_{s}+L_{p}\right)
	dt^{\prime}}\left\{  g(t^{\prime\prime})A(q)f(q,Q_{w})M\dot{Q}\right\}  .
\end{equation}
The functional between $\left\{  ..\right\}  $ appears as a weight to the
uncoupled path integral.\ This weight is only evaluated at $t=t^{\prime\prime
}$. Adapting the reasoning given by Feynman and Hibbs \cite{Feynman-Hibbs} to
our present problem, this is seen to be tantamount to uncoupled propagation
from $t_{1}=t_{w}-\tau/2$ to $t=t^{\prime\prime}$ and between $t^{\prime
	\prime}$ and $t_{2}=t_{w}+\tau/2$ while at $t=t^{\prime\prime}$ the term
$\left\{  g(t^{\prime\prime})A(q)f(q,Q_{w})M\dot{Q}\right\}  $ is evaluated at
each of the positions $q(t^{\prime\prime})$ compatible with the uncoupled paths.

We now use Eq. (\ref{Agdef}) as is usually done in weak measurements and as we
have done in Eq. (\ref{Ag}) for the classical pointer motion. This means that
the weak coupling of duration $\tau$ is seen as taking place at the average
time $t=t_{w}$ with an effective coupling $g$. We thereby obtain Eq.
(9) of the main text,%
\begin{align}
&  K_{int}(X_{2},x_{2},t_{2};X_{1},x_{1},t_{1})=K_{p}(X_{2},t_{2};X_{1}%
,t_{1})K_{s}(x_{2},t_{2};x_{1},t_{1})\nonumber\\
&  -\frac{ig}{\hbar}\int_{X_{1}}^{X_{2}}D[Q(t)]e^{\frac{i}{\hbar}\int_{t_{1}%
	}^{t_{2}}L_{p}dt^{\prime}}M\dot{Q}\int dqK_{s}(x_{2},t_{2};q,t_{w}%
)A(q)f(q,Q_{w})K_{s}(q,t_{w};x_{1},t_{1}).\label{Aintts}%
\end{align}
The term $\dot{Q}$ taken at $t_{w}$ should be understood here as the variation
$\left(  Q(t_{w}+\varepsilon)-Q(t_{w})\right)  /\varepsilon$ along each path,
where $\varepsilon$ is an \ infinitesimal time variation at $t_{w} $.

We now take into account the pre and postselected states and focus on the
probe evolution. The initial state
\begin{equation}
\left\vert \Psi_{i}\right\rangle =\int dX_{i}dx_{i}\left\vert X_{i}%
\right\rangle \left\vert x_{i}\right\rangle \psi(x_{i})\phi(X_{i})
\end{equation}
is propagated up to $t_{w}$ employing the standard uncoupled
propagators.\ Assuming no self-evolution for the probe, this can be written as%
\begin{equation}
\Psi(X,x,t_{w})=\psi(x,t_{w})\phi(X,t_{w})
\end{equation}
where $\psi(x,t_{w})=\int dx_{i}K_{s}(x,t_{w};x_{i},t_{i}))\psi(x_{i},t_{i}).
$ The postselected state $\left\langle b_{f}\right\vert =\int dxb_{f}^{\ast
}(x_{f})\left\langle x_{f}\right\vert $ is written as the propagated state of
the wavefunction $b_{f}^{\ast}(x,t_{w})$,%
\begin{equation}
b_{f}^{\ast}(x_{f},t_{f})=\int dxK_{s}^{\ast}(x_{f},t_{f};x,t_{w}))b_{f}%
^{\ast}(x,t_{w});
\end{equation}
$b_{f}^{\ast}(x,t_{w})$ appears as the postselected state propagated backward
in time from $t_{f}$ to $t_{w}$. Hence%
\begin{align}
\left\langle X_{f}\right\vert \left\langle b_{f}(t_{f})\right\vert \hat
{U}(t_{f},t_{i})\left\vert \Psi_{i}(t_{i})\right\rangle  &  =\int_{X}^{X_{f}%
}D[Q(t)]e^{\frac{i}{\hbar}\int_{t_{_{w}}}^{t_{f}}L_{p}dt^{\prime}}\left[  \int
dxdx_{f}K_{s}^{\ast}(x_{f},t_{f};x,t_{w}))b_{f}^{\ast}(x,t_{w})\psi
(x,t_{w})\right. \nonumber\\
&  \left.  -\frac{ig}{\hbar}M\dot{Q}\int dxdx_{f}K_{s}^{\ast}(x_{f}%
,t_{f};x,t_{w}))b_{f}^{\ast}(x,t_{w})A(q)f(q,Q_{w})\psi(x,t_{w})\right]
\phi(X,t_{w}).
\end{align}
We reexponentiate and obtain%
\begin{align}
\left\langle X_{f}\right\vert \left\langle b_{f}(t_{f})\right\vert \hat
{U}(t_{f},t_{i})\left\vert \Psi_{i}(t_{i})\right\rangle  &  =\int_{X}^{X_{f}%
}D[Q(t)]e^{\frac{i}{\hbar}\int_{t_{_{w}}}^{t_{f}}L_{p}dt^{\prime}}\int
dxdx_{f}K_{s}^{\ast}(x_{f},t_{f};x,t_{w}))b_{f}^{\ast}(x,t_{w})\psi
(x,t_{w})\nonumber\\
&  \times\exp\left[  -\frac{ig}{\hbar}M\dot{Q}\left(  \frac{\int dxdx_{f}%
	K_{s}^{\ast}(x_{f},t_{f};x,t_{w}))b_{f}^{\ast}(x,t_{w})A(q)f(q,Q_{w}%
	)\psi(x,t_{w})}{\int dxdx_{f}K_{s}^{\ast}(x_{f},t_{f};x,t_{w}))b_{f}^{\ast
	}(x,t_{w})\psi(x,t_{w})}\right)  \right]  \phi(X,t_{w}).\label{Aexp1}%
\end{align}
\ 

Eq. (\ref{Aexp1}) gives the probe state at time $t_{f}$ correlated with
postselection of state $\left\vert b_{f}\right\rangle $ and can be rewritten
as
\begin{align}
\phi_{b_{f}}(X_{f},t_{f})  &  =\int_{X}^{X_{f}}D[Q(t)]e^{\frac{i}{\hbar}%
	\int_{t_{_{w}}}^{t_{f}}L_{p}dt^{\prime}}\exp\left[  -\frac{ig}{\hbar}M\dot
{Q}A^{w}\right]  \phi(X,t_{w})\nonumber\\
&  \times\int dxdx_{f}K_{s}^{\ast}(x_{f},t_{f};x,t_{w}))b_{f}^{\ast}%
(x,t_{w})\psi(x,t_{w})\label{Aexp2}%
\end{align}
where $A^{w}$ is the weak value. Recalling that $\dot{Q}$ stands for $\left(
Q(t_{w}+\varepsilon)-Q(t_{w})\right)  /\varepsilon$ , Eq. (\ref{Aexp2}) indeed
shifts $\phi(X,t_{w})$ to $\phi(X-gA^{w},t_{w})$ in the interval $\left[
t_{w},t_{w}+\varepsilon\right]  $ and then propagates freely up to $t_{f}%
.$Note that the probe shift seems to happen at $t=t_{w}$ although the shift
depends on the postselection taking place at a future time $t_{f}.\ $This is
the result of reexponentiating a first order expression -- it must be clear
that Eqs. (\ref{Aexp1}) and (\ref{Aexp2}) are only valid at $t=t_{f} $.

\section*{Appendix 2 - The classical limit}

The semi-classical expansion of the path integral is well-known. We quote here
the generic result (see eg \cite{schulman} or \cite{handbook}) in which the
first order term in the $\hbar$ expansion of the propagator appears as a sum
over all the classical paths $k$ connecting the initial and end points in time
$t_{f}-t_{i}$. This term, denoted by $K^{sc}(x_{f},t_{f};x_{i},t_{i}) $ is
given by%
\begin{align}
K(x_{f},t_{f};x_{i},t_{i})  &  =\int_{x_{i}}^{x_{f}}\mathcal{D}[q(t)]\exp
\left(  \frac{i}{\hbar}\int_{t_{i}}^{t_{f}}\left(  \frac{m\dot{q}^{2}}%
{2}-V(q)\right)  dt^{\prime}\right) \\
&  \underset{\hbar/S\rightarrow0}{\simeq}\sum_{k}\mathcal{A}_{k}(x_{f}%
,t_{f};x_{i},t_{i})\exp i\left[  \mathcal{S}_{k}(x_{f},t_{f};x_{i}%
,t_{i})/\hbar-\pi\mu_{k}/2\right] \label{ksc}%
\end{align}
where $\mathcal{S}_{k}(x_{f},t_{f};x_{i},t_{i})$ is the action of the
classical trajectory $k$ while $\mathcal{A}_{k}(x_{f},t_{f};x_{i},t_{i})$ is
the semi-classical amplitude of the classical trajectory $k$ given by
\begin{equation}
\mathcal{A}_{k}(x_{f},t_{f};x_{i},t_{i})=\left(  \frac{1}{2\pi i\hbar}\right)
^{d/2}\left[  \det\left(  -\frac{\partial_{k}^{2}\mathcal{S}}{\partial
	x_{f}^{m}\partial x_{i}^{n}}\right)  \right]  ^{1/2}.
\end{equation}
$d$ is the spatial dimension, so that $x$ has coordinates $x=(x^{1}%
,x^{2},...,x^{d}).$ $\mu_{k}$, counts the number of conjugate points on the
trajectory $k$; as in the main text we absorb $\mu_{k}$ into the action to
simplify the notation.

The squared modulus $\left\vert \mathcal{A}_{k}(x_{f},t_{f};x_{i}%
,t_{i})\right\vert ^{2}$ obeys the classical transport equation
\cite{littlejohn}.\ It is a purely classical quantity (leaving aside the
normalizing prefactor) that can be obtained by solving the equations of motion
from the Lagrangian (\ref{Alag}). However a semiclassically propagated
wavefunction, $\zeta(x_{f},t_{f})=\int dx_{i}K^{sc}(x_{f},t_{f};x_{i}%
,t_{i})\zeta(x_{i},t_{i})$ remains a quantum object: the wavefunction evolves
through a coherent superposition of interfering waves carried by classical
paths. Classical distributions appear when the density matrices (here for the
pure state $\left\vert \zeta\right\rangle $, $\rho(x^{\prime},x,t)=\zeta
^{\ast}(x^{\prime},t)\zeta(x,t)$) become diagonal. It can then be shown (eg,
\cite{bhatt}) that $\rho(x,x,t)$ obeys the classical Liouville equation for
the configuration space probability distributions. Heuristically, the idea is
that the propagator for the density matrix,%
\begin{equation}
\sum_{kk^{\prime}}\mathcal{A}_{k^{\prime}}^{\ast}(x^{\prime},t;x_{i}^{\prime
},t_{i})\mathcal{A}_{k}(x,t;x_{i},t_{i})\exp i\left[  \mathcal{S}%
_{k}(x,t;x_{i},t_{i})/\hbar-\mathcal{S}_{k^{\prime}}(x^{\prime},t^{\prime
};x_{i}^{\prime},t_{i})/\hbar\right]
\end{equation}
becomes diagonal when coarse-graining over a classical scale , given that the
exponential oscillates wildly (since $\mathcal{S}_{k}/\hbar\rightarrow\infty$)
and cancels out when averaged over a classical scale \cite{classlim}. \ Hence
only the diagonal term $\sum_{k}\left\vert \mathcal{A}_{k}(x,t;x_{i}%
,t_{i})\right\vert ^{2}$ survives -- and this is precisely the classical
density of paths obeying the classical transport equation.

In the present problem $K^{sc}(X_{f},x_{f},t_{f};X_{i},x_{i},t_{i})$
represents the semiclassical propagator for the coupled system-probe
evolution, Eq.
(7). Since the actions are large, the weak coupling term cannot be expanded to
first order. Hence the density matrix $\rho(t_{i})=\left\vert \Psi
(t_{i})\right\rangle \left\langle \Psi(t_{i})\right\vert $ evolves according
to $\rho(t)=\hat{U}(t,t_{i})\rho(t_{i})\hat{U}^{\dagger}(t,t_{i})$, where the
evolution operators $U$ are written in terms of $K^{sc}$. Coarse-graining
leads to keeping only the diagonal components $\rho(X,x)=\sum_{J}\left\vert
\mathcal{A}_{J}^{s}(X,X^{\prime})\right\vert ^{2}\left\vert \mathcal{A}%
_{J}^{p}(x,x^{\prime})\right\vert ^{2}$. The coarse-grained initial density
matrix of the system becomes a classical distribution $\rho_{s}(q,t_{i})$ that
evolves according to the Liouville equation.\ Put differently, each point in
configuration space of the initial distribution evolves according to the
classical equations of motion obtained from the Lagrangian (\ref{Alag}) so
that each point of the initial distribution leads to a probe shift obtained
from Eq. (\ref{Ag}):%
\begin{equation}
\Delta Q\equiv Q(t>t_{w}+\tau/2)-Q(t<t_{w}-\tau/2)=\int dt^{\prime}%
g(t^{\prime})A(q)f(q,Q_{w}).
\end{equation}
The average pointer shift is obtained by interating over the system
distribution: denoting by $\rho_{s}(q,t_{i})$ the initial normalized classical
distribution (for instance the coarse-grained pre-selected system density
matrix), $\rho_{s}(q,t_{i})$ evolves to $\rho_{s}(q,t_{w})$ by the time the
interaction takes place and the average shift is obtained by integrating over
the distribution,%
\begin{equation}
\overline{\Delta Q}=g\int_{{}}A(q,t_{w})f(q,Q_{w})\rho_{s}(q,t_{w})dq.
\end{equation}

We finally need to take into account postselection. From a classical
viewpoint, postselection is a filter that puts a condition on a chosen
configuration space function $b(q,t_{f})$ at some subsequent time $t_{f}.$ Let
us denote this condition by $\mathfrak{B}_{f}$, so that postselection is
successful whenever $b(q,t_{f})\in\mathfrak{B}_{f}$. The average shift
conditioned on $b(q,t_{f})\in\mathfrak{B}_{f}$ can now be computed by taking
into account the fraction of the system distribution $\rho_{s}(q,t_{w})$ that
will end up in $\mathfrak{B}_{f}$ at time $t_{f}.$ By evolving backward the
classical equation of motion, this distribution can be written as
\begin{equation}
\int_{\mathfrak{B}_{w}}\rho_{s}(q^{\prime},t_{w})dq^{\prime}%
\end{equation}
where the domain of integration $\mathfrak{B}_{w}$ (at $t_{w}$) contains all
the configuration space points such that $b(q,t_{f})\in\mathfrak{B}_{f}.$ The
average probe shift conditioned on $b(q,t_{f})\in\mathfrak{B}_{f}$ is
therefore given by Eq.
(14):
\begin{equation}
\left\langle \Delta Q\right\rangle =g\int_{\mathfrak{B}_{w}}A(q,t_{w}%
)f(q,Q_{w})\frac{\rho_{s}(q,t_{w})}{\int_{\mathfrak{B}_{w}}\rho_{s}(q^{\prime
	},t_{w})dq^{\prime}}dq,\label{Aclassps}%
\end{equation}
where the denominator renormalizes $\rho_{s}$ over the postselected ensemble.
In the classical limit, $\left\langle \Delta Q\right\rangle $ is a standard
conditional expectation value that can never be anomalous.

\end{document}